\documentclass[aip,rsi,preprint]{revtex4-1}

\usepackage{graphicx}
\usepackage[squaren]{SIunits}

\begin{document}

\title{Three-dimensional quantum photonic elements based on single nitrogen
vacancy-centres in laser-written microstructures}

\author{Andreas W. Schell}
\email[Electronic mail: ]{andreas.schell@physik.hu-berlin.de}
\affiliation{Nano-Optics, Institute of Physics, Humboldt-Universit\"{a}t zu
Berlin, Newtonstra{\ss}e~15, D-12489 Berlin, Germany}

\author{Johannes Kaschke}
\affiliation{Institute of Applied Physics,
DFG-Center for Functional Nanostructures,
Institute of Nanotechnology,
Karlsruhe Institute of Technology (KIT),
76128 Karlsruhe, Germany}

\author{Joachim Fischer}
\affiliation{Institute of Applied Physics,
DFG-Center for Functional Nanostructures,
Institute of Nanotechnology,
Karlsruhe Institute of Technology (KIT),
76128 Karlsruhe, Germany}

\author{Rico Henze}
\affiliation{Nano-Optics, Institute of Physics, Humboldt-Universit\"{a}t zu
Berlin, Newtonstra{\ss}e~15, D-12489 Berlin, Germany}

\author{Janik Wolters}
\affiliation{Nano-Optics, Institute of Physics, Humboldt-Universit\"{a}t zu
Berlin, Newtonstra{\ss}e~15, D-12489 Berlin, Germany}

\author{Martin Wegener}
\affiliation{Institute of Applied Physics,
DFG-Center for Functional Nanostructures,
Institute of Nanotechnology,
Karlsruhe Institute of Technology (KIT),
76128 Karlsruhe, Germany}

\author{Oliver Benson}
\affiliation{Nano-Optics, Institute of Physics, Humboldt-Universit\"{a}t zu
Berlin, Newtonstra{\ss}e~15, D-12489 Berlin, Germany}

\begin{abstract}

To fully integrate quantum optical technology, active quantum systems must be combined with resonant
microstructures and optical interconnects harvesting and routing photons in three diemsnsions (3D) on one chip. 
We fabricate such combined structures for the first time by
using two-photon laser lithography and a photoresist containing nanodiamonds including nitrogen vacancy-centers. 
As an example for possible functionality, single-photon generation, collection, and transport 
is successfully accomplished.
The single photons are efficiently collected via resonators and routed in 3D through waveguides, all on one 
optical chip. Our one-step fabrication scheme is easy to implement,
scalable and flexible. Thus, other complex assemblies of 3D quantum optical structures are feasible as well.

\end{abstract}

\maketitle

A fully integrated quantum optical technology requires active quantum systems incorporated into resonant
optical microstructures and inter-connected in three dimensions via photonic wires. Nitrogen vacancy-centres
(NV-centres) in diamond which are excellent photostable room temperature single-photon emitters
\cite{Kurtsiefer2000} are ideal candidates for that purpose \cite{Jelezko2006, Aharonovich2011}.
Extensive research efforts to couple NV-centres to photonic structures such as optical microresonators
\cite{Barclay2009,Hausmann2012}, microcavities \cite{Wolters2010,Sar2011,Riedrich-Moller2012}, and
waveguides \cite{Fu2008,Schroder2010,Schroder2012} have been pursued. Strategies for integration range from
top-down fabrication via etching of diamond membranes \cite{Riedrich-Moller2012,Hausmann2012} to sophisticated
bottom-up assembly of hybrid structures using diamond nanocrystals
\cite{Wolters2010,Sar2011,Barclay2009} where the latter approach allows for deterministic coupling.
Recently, another approach based on the incorporation of nanodiamonds in soft glass optical fibres via a melting
process has been introduced \cite{Henderson2011}. Here, we utilize two-photon direct laser writing (DLW) to
fabricate fully three-dimensional (3D) structures from a photoresist mixed with a solution of nanodiamonds
containing NV-centres. For the first time, this approach facilitates building integrated 3D quantum photonic
elements of nearly arbitrary shapes.

\begin{figure*}
  \includegraphics{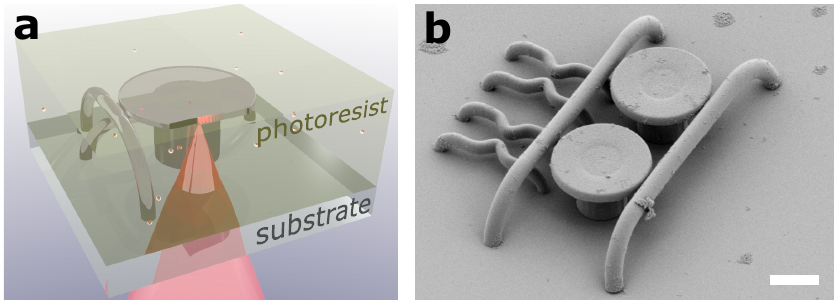}
  \caption{\textbf{Direct laser writing in nanodiamond photoresist.} \textbf{a}, Sketch of the direct laser writing process.
  A femtosecond laser beam is focussed into the photoresist in order to polymerize well defined 3D structures. \textbf{b}, Scanning electron
  micrograph of such a structure after development containing several key photonic elements, such as waveguides, couplers and microdisc resonators.
  Scale bar is \unit{5}{\micro\metre}.}
  \label{fig:dlw}
\end{figure*}

On-chip photonic circuits working at the single-quantum level play an important role for future quantum
information processing \cite{O'Brien2009}. Several
approaches to assemble such quantum photonic devices from different fundamental photonic entities have been
pursued \cite{Benson2011}. For example, by using self-assembled quantum dots in
semiconductor membranes, one can exploit the full power of semiconductor nanofabrication technology, and
sophisticated structures have been demonstrated \cite{O'Brien2009}, even with on-demand coupling 
architectures \cite{Badolato2005}. However, this approach is limited to two-dimensional structures, except for few results obtained by
extremely challenging manual membrane-stacking \cite{Aoki2008}.

Another easy and low-cost way of fabricating photonic structures is optical lithography via direct laser writing
(DLW) \cite{kawata2001finer,deubel2004direct} where a tightly focussed femtosecond laser beam is
used to expose a photoresist (see Figure \ref{fig:dlw}a). The use of multi-photon absorption enables a
sequential 3D exposure by scanning the sample or the focus of the laser. For common negative-tone photoresists,
unexposed parts are removed during a development step and the 3D polymer structures remain (see Figure
\ref{fig:dlw}b). DLW is well known for the fabrication of photonic crystals \cite{deubel2004direct}
or other photonic elements like resonators  \cite{liu2010direct,grossmann2011direct} and waveguides
\cite{Lee2012}. In order to functionalise the structures with optically active material, fluorescent dyes
\cite{sun2001two}, quantum dots \cite{Li2006} and
metal nanoparticles \cite{Shukla2011} have been incorporated. However, until
today there has been no 3D structure operating at the fundamental quantum level with single photons from single emitters
being collected and routed. Moreover, no combinations of multiple optical elements (different resonators,
couplers, waveguides) have been demonstrated. The reason is the lack of photostable quantum emitters which are
compatible with the DLW process while still preserving the possibility for high-quality DLW fabrication.

\subsection*{Results}

Here, we overcome these obstacles by mixing small amounts of nanodiamonds from solution directly into a known 
photoresist \cite{Fischer2011} (See Methods). This special photoresist allows for the fabrication
of transparent optical elements made from an acrylate polymer containing nanodiamonds. The nanodiamonds contain
NV-centres which are photostable even after exposure by the DLW laser. After the DLW process they serve as
integrated single-photon sources. It is especially noteworthy that our technique is not limited to NV-centres in
diamond nanocrystals. It can be applied to any photostable single-photon emitter in nanocrystalline material,
e.g. other diamond defect centres, such as Silicon vacancy-centres \cite{Neu2011}, or other deep defect centres
 in large band gap material \cite{Weber2010}.

In the following, we demonstrate the fabrication, functionality, and interaction of key building blocks for quantum photonic
circuitry. These are interconnects (i.e., waveguides) and functional elements like resonators and emitters. We
measure single-photon emission from a diamond nanocrystal in a waveguide-coupled resonator which demonstrates
the functional combination of all these building blocks and paves the way to more complex devices.

\begin{figure*}
  \includegraphics{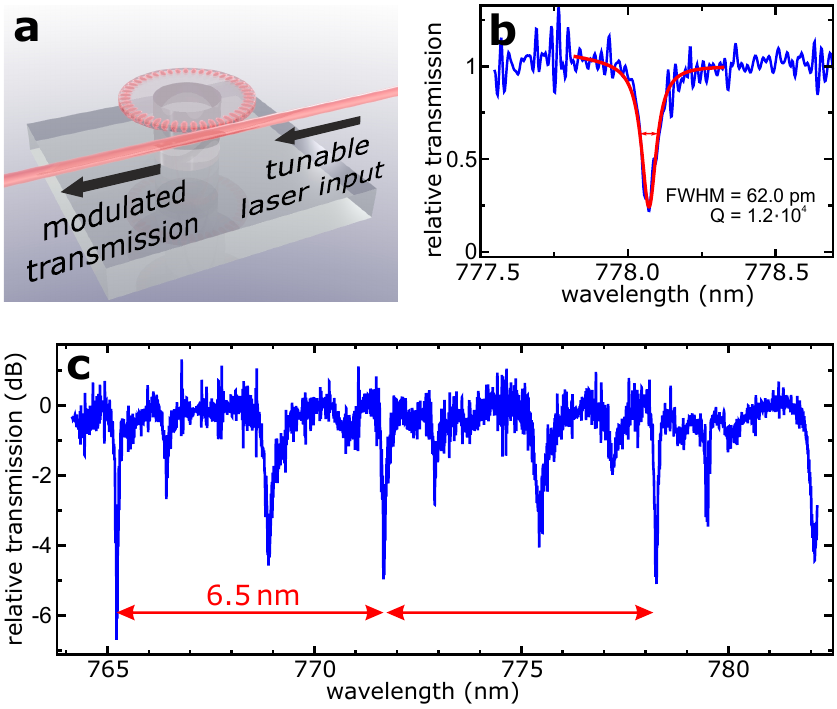}
  \caption{\textbf{Mode measurements}. \textbf{a}, A tunable laser is coupled to a disc resonator via a tapered fibre. Upon sweeping the
  laser frequency, the transmitted light is modulated by the modes of the resonator. \textbf{b}, Resonator mode with a quality factor Q of
  $1.2 \times 10^4$. \textbf{c}, Scan over many modes with a free spectral range of \unit{6.5}{\nano\meter}
  indicated by the red arrows.}
  \label{fig:tapercoupling}
\end{figure*}

As a first component for integrated photonic circuits, disc resonators with a disc diameter of
\unit{20}{\micro\metre} and a disc thickness of approximately \unit{1.2}{\micro\metre} on a stem with diameter
 \unit{10}{\micro\metre} were fabricated using the photoresist functionalised with nanocrystals (see Methods).
The resonator supports whispering gallery modes which are confined in the
disc close to the rim with high quality factors (Q-factors). 
We measured the Q-factor using a tunable laser coupled into the resonator via a tapered
fibre\cite{Vahala2003}. We found Q-factors as large as $10^4$ at a wavelength of around \unit{770}{\nano\metre} (see Figure
\ref{fig:tapercoupling} and Methods). The observed free spectral range of \unit{6.5}{\nano\meter} of the
resonator corresponds well to the expected value of \unit{6.4}{\nano\meter} for a disc with diameter \unit{20}{\micro\meter}
derived from geometrical considerations assuming an index of refraction of 1.5.

\begin{figure*}
  \includegraphics{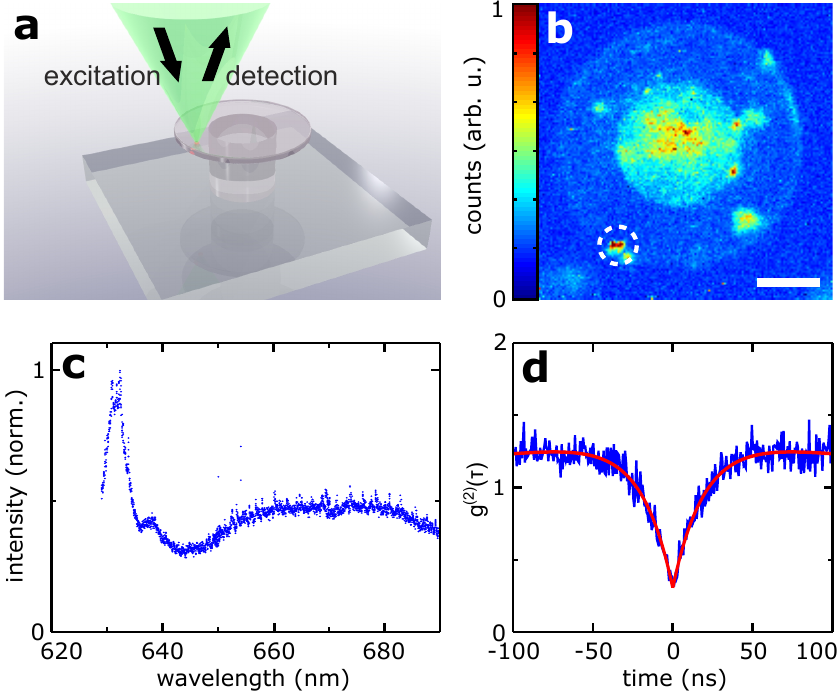}
  \caption{\textbf{DLW resonator containing single NV-centres.} A diamond nanocrystal containing a single NV-centre is coupled to whispering gallery modes
  of a DLW disc resonator. \textbf{a}, Measurement scheme. Detection and excitation take place at the same point in a confocal configuration.
  \textbf{b}, Scanning confocal image of the resonator disc. The circle indicates a bright spot identified as single NV-centre. Its
  fluorescence is analysed in \textbf{c} and \textbf{d}. Scale bar is \unit{5}{\micro\metre}. \textbf{c},
Spectrum of the collected fluorescence. The resonator modes are seen on the broad NV-centre phonon sidebands. The
peak at \unit{630}{\nano\meter} stems from the photoresist and can be bleached over time.
  \textbf{d}, Autocorrelation function of the fluorescence from the NV-centre. A clear antibunching behaviour can be seen. The red curve is a fit to the data.}
  \label{fig:diainresonator}
\end{figure*}

Next, a homebuilt confocal microscope (see Methods) was used to raster scan the fabricated resonator discs  (see Figure 3a) and
to identify single NV-centres by measuring the second-order autocorrelation function $g^{(2)}(\tau)$ in a
Hanbury Brown and Twiss (HBT) interferometer. In parallel, a grating spectrometer was used to resolve the
emission spectra of the individual emitters. Figure \ref{fig:diainresonator}b shows a confocal scan of a
resonator where fluorescent defects can be identified as bright spots on the resonator. Encircled is a spot on
the resonator's outer rim where coupling to the disc's whispering gallery modes is expected. Figure
\ref{fig:diainresonator}d displays an autocorrelation measurement of fluorescence collected from that spot. A
clear antibunching dip is observed reaching a value of $g^{(2)}(0)=0.31\pm0.04$ (standard error) as deduced from the fit shown as red
curve. No background correction was applied to these data. This measurement shows that the bright spot indeed
corresponds to a single NV-centre in the disc resonator.

The fluorescence spectrum of an NV-centre at room temperature is typically broadened over \unit{200}{\nano\meter} from approx.
\unit{600}{\nano\meter} to \unit{800}{\nano\meter} by phonon sidebands. This corresponds well to the measured emission spectrum shown in Figure
\ref{fig:diainresonator}c. The peak at approximately \unit{630}{\nano\meter} stems from fluorescence of the
photoresist and bleaches after long excitation. For confocal scans and correlation measurements this background
light was suppressed by spectral filtering. In addition, there is a fine modulation of the spectrum due to the
cavity resonances. It is attributed to photons that were initially emitted into the resonator modes and
afterwards scattered out by the diamond nanocrystal. Hence, these photons are detected in addition to the flat
unmodulated spectrum emitted directly out of the disc.

\begin{figure*}
  \includegraphics{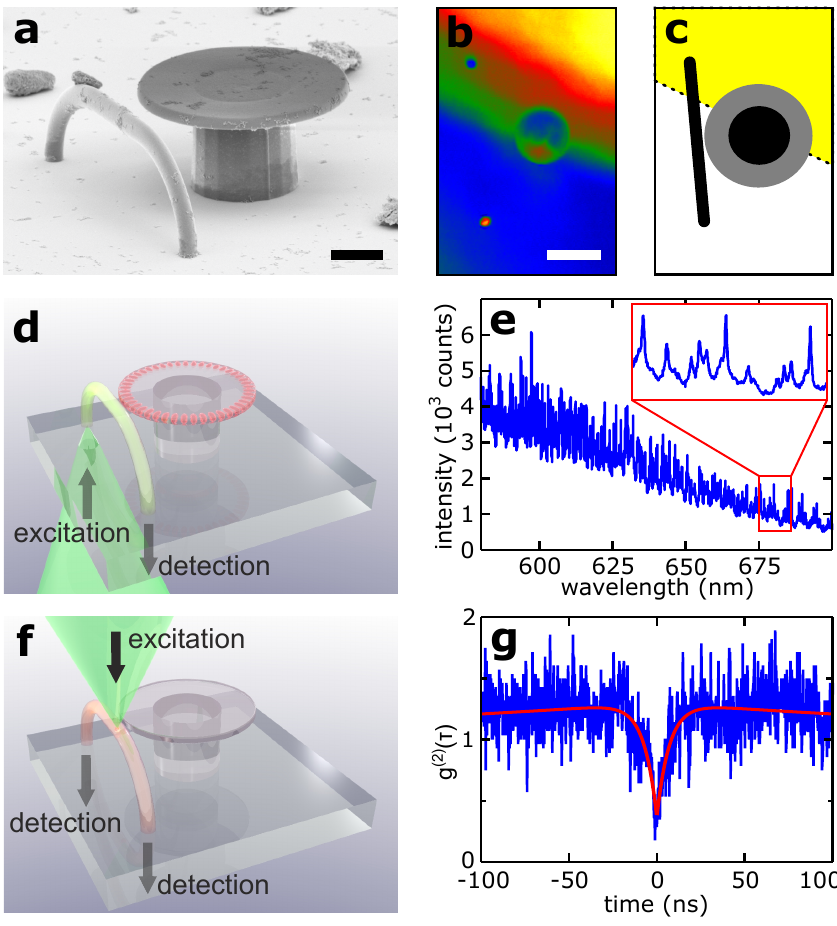}
  \caption{\textbf{Arc waveguide coupled to resonator.}
  \textbf{a}, Scanning electron micrograph of a 3D arc waveguide/resonator assembly. Length of the scale bar is \unit{5}{\micro\metre}.
  \textbf{b}, The waveguide resonator assembly as seen through the cover slip. The upper half of the image is illuminated from below by white light.
  Efficient light guiding through the arc waveguide can be seen. Scale bar is \unit{10}{\micro\metre}. \textbf{c}, Sketch of the
  region shown in \textbf{b} with the illuminated area marked yellow. \textbf{d}, Sketch how the spectrum in \textbf{e} was acquired.
  \textbf{e}, Spectrum of the fluorescence detected at one of the waveguide's ports while exciting through the other.
  Resonator modes can be clearly identified. \textbf{f}, Experimental configuration for the autocorrelation measurement in \textbf{g}.
  A second objective is used for excitation.
  \textbf{g}, Measured autocorrelation function of a NV-centre inside a waveguide. It is acquired at both output ports of the waveguide
  in a cross-correlation configuration.}
  \label{fig:waveguide}
\end{figure*}

As a second photonic element, arc waveguides  from the same material with a width of \unit{1.8}{\micro\meter} and
a length of \unit{40}{\micro\meter} were built. The waveguides were fabricated close to the disc resonator
in order to enable evanescent coupling to the resonator's modes (see Figure \ref{fig:waveguide}a). Their ability
to guide light within a broad spectral range is demonstrated by illuminating one end of the waveguide with light from a halogen lamp as seen
in Figures \ref{fig:waveguide}b and c. The blue and red circular spots correspond to the input and output of
the arc waveguide, respectively.

Coupling via the waveguides to the resonator modes was confirmed by shining an excitation laser (wavelength of
\unit{532}{\nano\metre}) into one port of the waveguide and, after blocking excitation light by a dichroic
mirror, analysing the light coming out of the other port. Obviously, the \unit{532}{\nano\metre}-laser excites
background fluorescence in the resonator, which is then coupled back into the waveguide. The measured
fluorescence spectrum in Figure \ref{fig:waveguide}e thus shows the characteristic modulation by resonator
modes and proves a good coupling between guided modes in the arc waveguide and confined modes in the resonator
(see Supplementary Information).

By moving the excitation spot of a second objective over a waveguide (see Figure \ref{fig:waveguide}a), individual nanodiamonds inside 
can be addressed and identified as spots with higher fluorescence. The intensity cross correlation
between the waveguide's two outputs shows a clear antibunching behaviour. This result highlights a way of
coupling single emitters to thin waveguides, which can possess very high collection efficiencies, 
as demonstrated recently using  tapered optical fibres
\cite{Fujiwara2011,Schroder2012}. Here, we show a direct on-chip integration of such devices with both waveguide
ends accessible. Detection at the waveguide's ends can yield count rates of 
 \unit{29,000}{\per \second} after passing a 50/50 beam splitter and spatial and spectral
filtering. The autocorrelation function at zero time delay of the fluorescence collected in this way is
$g^{(2)}(0)=0.37\pm0.13$ (standard error) showing the single-photon character of the light (see Figure \ref{fig:waveguide}g). When
accounting for the 50/50 beam splitter, at high excitation powers the number of background corrected counts
from the defect centre is measured to be \unit{130,000}{\per \second}.

\begin{figure*}
  \includegraphics{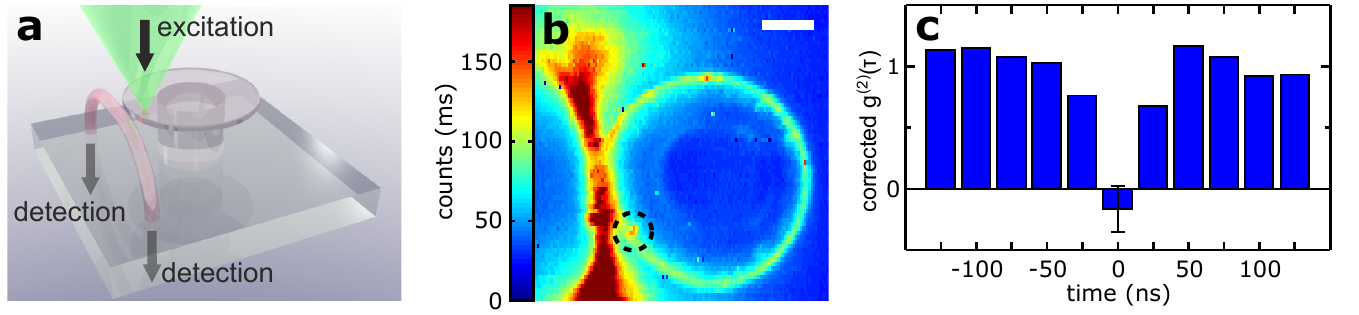}
  \caption{\textbf{Waveguide coupling of a single NV-centre inside a resonator.} \textbf{a}, Sketch of the experimental configuration.
  The excitation spot is scanned over the resonator disc. Photons are detected at both waveguide outputs simultaneously.
  \textbf{b}, Photon counts collected at
  one end of the waveguide while scanning the excitation spot with a second objective. The circle
  highlights the position of a single NV-centre. Shape distortions are due to non closed-loop piezo-scanning. Scale bar is \unit{5}{\micro\metre}.
  \textbf{c}, Cross correlation (background corrected) between the waveguide output ports when the NV-centre marked by a
  circle in \textbf{b} is excited. The bin size corresponds to the repetition rate of \unit{40}{\mega\hertz}}
  \label{fig:res16dia}
\end{figure*}

Finally, we verified active quantum functionality of a coupled 3D system, i.e., single photon emission to
resonator modes, as well as subsequent collection and routing via a coupled waveguide. In order to do so, raster
scans of the excitation spot over the resonator disc were performed while the photons from one of the
waveguide's ends are collected (see Figure \ref{fig:res16dia}a). When the excitation laser hits the outer rim of
the resonator, the created fluorescence light is coupled to the resonator modes and subsequently out to the
waveguide, as can bee seen in Figure \ref{fig:res16dia}b. Therefore, the outer rim of the resonator appears
bright in the raster image. Bright spots indicate possible positions of NV-centres. To identify NV-centres via
their non-classical light emission, an HBT setup, with the laser-written arc waveguide serving as beam splitter, is
used.
Figure \ref{fig:res16dia}b shows this correlation measurement for the NV-centre encircled in Figure
\ref{fig:res16dia}a. The data shown were acquired with a pulsed laser at \unit{40}{\mega\hertz} repetition rate and background
corrected (see Supplementary Information). The value of $g^{(2)}(0)=-0.18\pm0.21$ (standard deviation) is far below the threshold 
$g^{(2)}(0)=0.5$ for having 
equal contributions of two centers and proves that the main contribution
to the signal stems from a single NV-centre. This clearly indicates that the NV-centre emits single photons
into resonator modes, which are then coupled out and redirected by the arc waveguide. With a single source in a resonator
coupled to a waveguide, this device represents a key integrated 3D quantum photonic circuit.

\subsection*{Discussion}

In conclusion, we have shown an approach which allows for the direct incorporation of single quantum emitters
into true 3D photonic structures of nearly arbitrary shape. The one step fabrication scheme is easy to implement
and requires no elaborate clean-room environment. We have pointed out that our method is not limited to
NV-centres in nanodiamond, but can be extended to any other stable quantum emitter embedded in nanocrystals. The
flexibility of DLW enables immediate scaling-up to more complex structures (see Figure \ref{fig:dlw}b) where
single photons are collected, sent through beam splitters, interferometers, or other optical elements
inter-connected by waveguides in a 3D architecture. 
Also tree-dimensional multi-arm quantum interferometers~\cite{} can be implemented.
Tuning of individual components, e.g. the resonator, is 
possible as well (see Supplementary Information). With (partial) metallisation of the structures 
embedding single photon emitters in 3D plasmonic structures or in metamaterials \cite{Rill2008} can be envisioned. 
A next challenging, but rewarding step is to perform DLW of perfectly aligned 3D structures around emitters of known position. 
This would enable on-demand fabrication of arbitrary 3D quantum photonic architectures.

\subsection*{Methods}

\textbf{Direct laser writing.} The DLW setup consists of a Ti:sapphire oscillator (SpectraPhysics MaiTai HP)
delivering \unit{100}{\femto\second} pulses centred around \unit{810}{\nano\metre}. The beam is focussed through
an oil-immersion lens (Leica HCX PL APO 100x/1.40-0.70 OIL CS). The sample is moved with a 3D piezo stage (PI
P-527.3CL). The writing was performed with \unit{6}{\milli\watt} average writing power and at
\unit{50}{\micro\metre\per\second} scan velocity. The samples were developed in isopropyl alcohol for 30
minutes, afterwards in acetone for 30 minutes, and finally dried super-critically (Leica CPD030).

The new photoresist is based on the monomer pentaerythritol tetraacrylate (PETTA) which contains 350~ppm
monomethyl ether hydroquinone as inhibitor. Next, 0.25~\% wt of the photoinitiator
7-diethylamino-3-thenoylcoumarin were added. Last, 2~\% wt of an ethanol-based nanodiamond suspension were added
and stirred overnight. Nanodiamonds were of type 1b and had a median diameter of \unit{25}{\nano\metre} (Microdiamant AG).
Spatial dimensions of the fabricated structures were extracted from scanning electron micrographs.

\textbf{Confocal microscope.} The homebuilt confocal microscope is based on the body of a Zeiss Axiovert 200.
Either a frequency doubled Nd:YAG continuous wave laser or a pulsed laser (LDH-P-FA-530, PicoQuant) were used to
create the excitation light around \unit{532}{\nano\metre}. A large-numerical-aperture air microscope objective
mounted on a 1D piezo stage (MIPOS 100, piezosystem jena) focussed the excitation light and collected the
fluorescence. The sample was scanned by a 2D piezo stage (PXY 80 D12, piezosystem jena). Collected fluorescence
light was separated from the excitation laser by a dichroic mirror. Then, it was either spatially filtered and
sent
to a grating spectrometer 
or it was spectrally filtered by RG630 glass, split by a 50/50 beam splitter, and directed towards two pinholes.
These pinholes could be independently adjusted to spatially filter out different spots of the image plane, so
that cross correlations could be measured. After the pinholes the light was directed to avalanche photodiodes
(SPCM-AQR-14/SPCM-AQRH-14, Perkin Elmer).

To perform correlation measurements at the waveguide ends, an oil-immersion microscope objective (UPlanSApo
60XO, Olympus) was used for collecting the light, while different air objectives were used to focus the
excitation laser onto the sample from the other side. The air objectives were mounted on a 3D piezo system
(PXY100 ID and MIPOS 100, piezosystem jena) to enable scanning of the excitation spot while the sample was kept
at a fixed position. Also, the 50/50 beam splitter within the HBT setup could be replaced by a D-shaped mirror
in an image plane in cases where cross-correlation measurements with high count rates were needed.

\textbf{Q-factor measurements.} The mode structure of the disc resonators was analysed by means of coupling in a
tunable external-cavity diode laser (Velocity Series, New Focus) at around \unit{770}{\nano\metre} via the
evanescent fields of a tapered optical fibre.
By tuning the frequency of the laser over the distinct whispering gallery modes, light was coupled into the
resonator and different modes could be observed as Lorentzian shaped dips in the transmitted power. The
polarization of the incoming light was chosen to maximize the coupling depth. For normalizing the data sets, we
first performed a reference scan with an uncoupled fibre taper and compared the results to the situation when
the same taper was coupled to the resonator. This provided direct access to the different coupling depths and
cancelled out overlaying power modulations caused by slightly non-adiabatic taper transitions.

The Q-factors were calculated from the dips by a Lorentzian fit function with an additional linear term to
better match the local environment of the resonances. The fibre taper used had a waist diameter of
\unit{1.5}{\micro\metre}. Measurements were performed in full contact.

\subsection*{Acknowledgements}
The Berlin team acknowledges support by DFG (FOR1493 and SFB951).
The Karlsruhe team acknowledges support by the
DFG-Center for Functional Nanostructures (CFN) via
subprojects A1.4 and A1.5 and by the Karlsruhe School of Optics \& Photonics (KSOP).
J.W. acknowledges funding by the state of Berlin (Elsa-Neumann). 

\subsection*{Author contributions}
A.W.S. wrote the manuscript and conceived the experiment with J.F., J.K. and J.F. produced the samples, A.W.S., R.H. and J.W.
performed the optical measurements and data analysis. M.W. and O.B. supervised the study.
All authors discussed the experiment, the results and the manuscript.

\subsection*{Additional Information}
The authors declare no competing financial interests.

\clearpage

\section*{Spectral mode analysis}

\begin{figure*}[b]
  \includegraphics[width=0.75\textwidth]{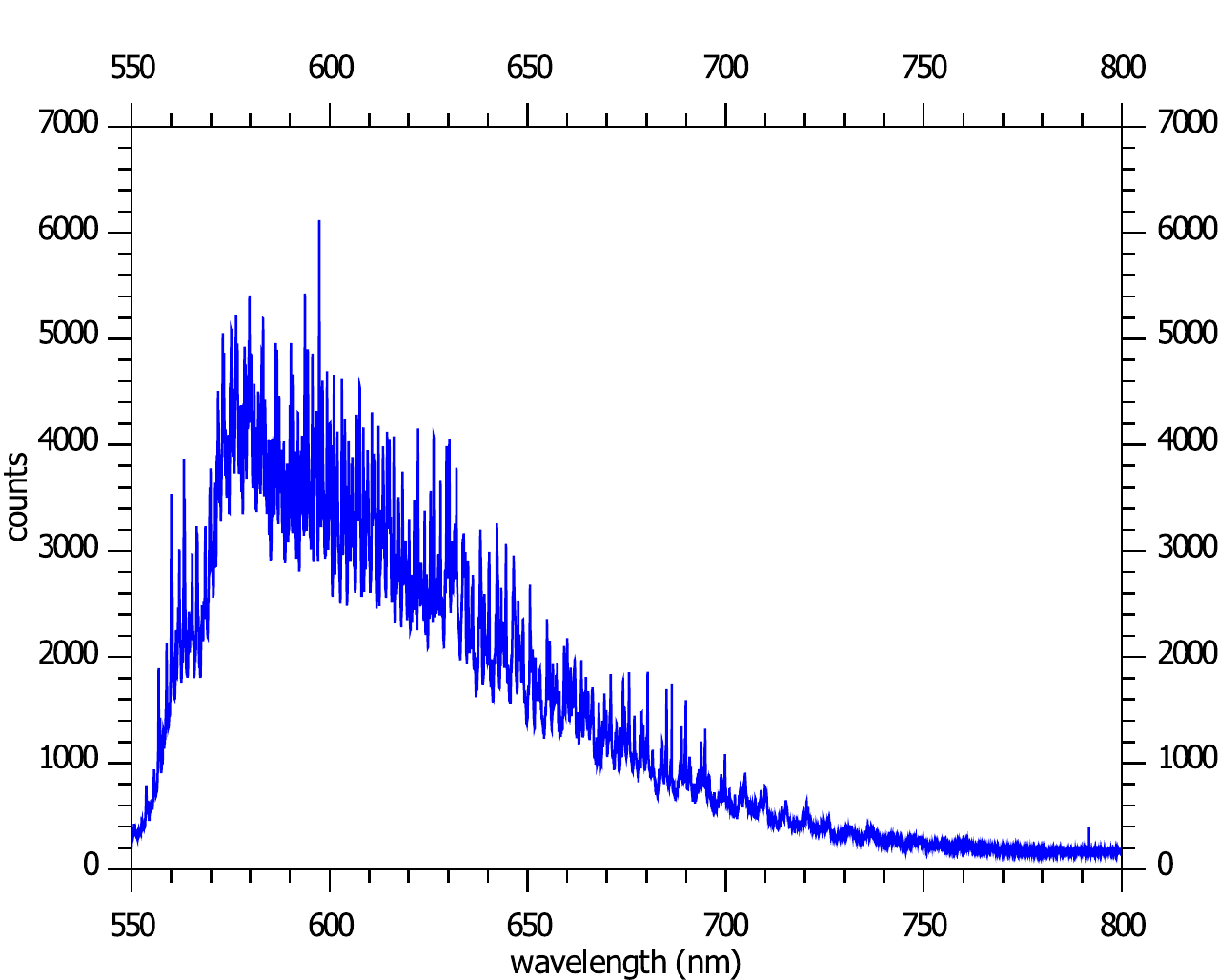}
  \caption{Spectrum of the background fluorescence when excited through one waveguide end and collected on the other (see main text)}  
  \label{fig:spec}
\end{figure*}

\begin{figure*}
  \includegraphics[width=0.75\textwidth]{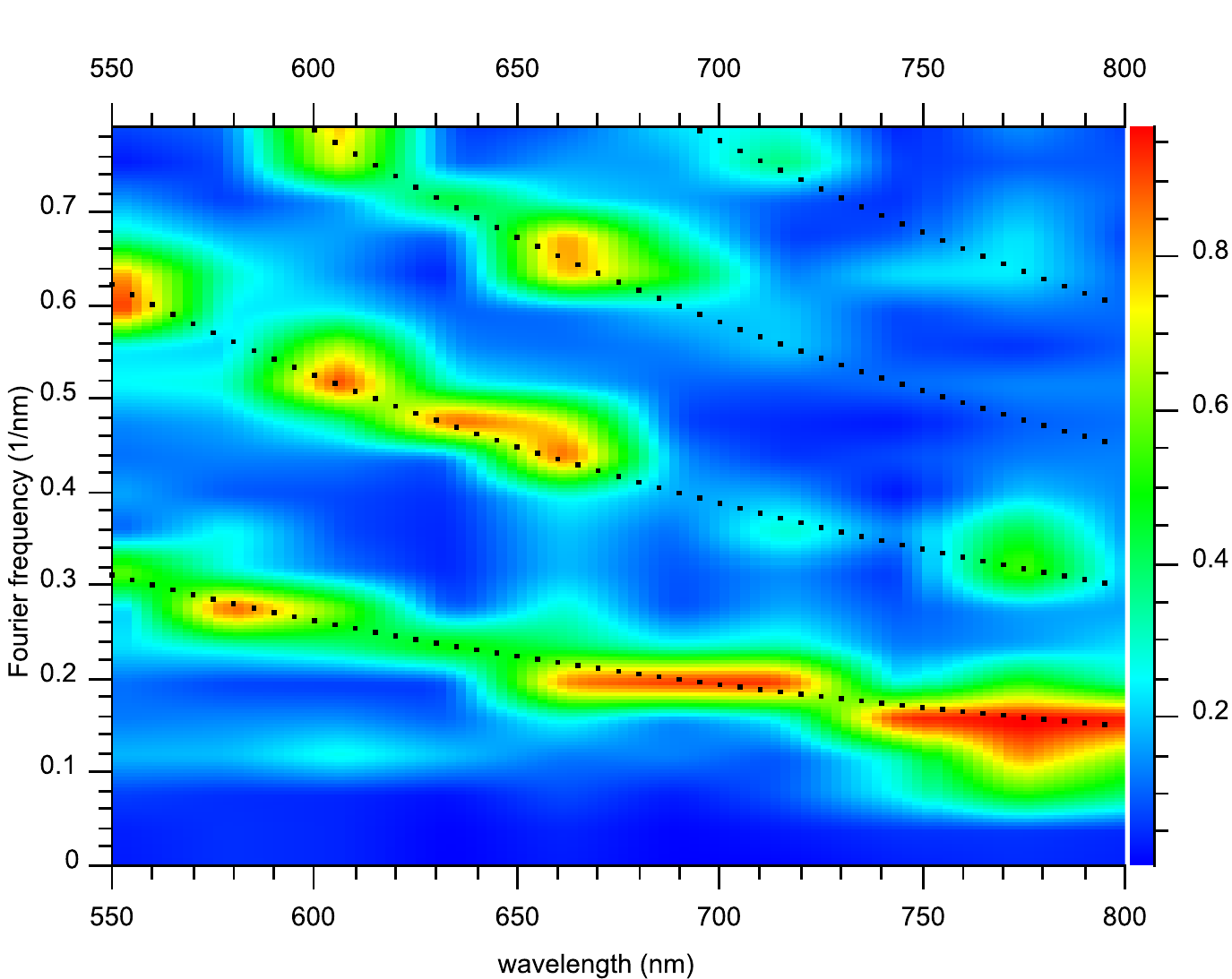}
  \caption{Normalised segmented Fourier transform showing the mode structure of the resonator's WGMs.
   The black dotted lines represent the different Fourier components of the collected spectrum. 
   They are fits to the expected Fourier frequencies of the resonator's modes and connected to 
   the wavelength dependent FSR of the system. The lowest frequency component directly relates to 
   1/FSR. Within the determined spectral window a constant refractive index of around 1.5 was 
   assumed for resonators with \unit{20}{\micro\meter} diameter.}  
  \label{fig:segfu}
\end{figure*}

For a quantitative analysis of the spectrum measured from the background fluorescence by a grating spectrometer 
(see Figure \ref{fig:spec}) we performed a segmented Fourier transformation on the data. We divided the set into \unit{25}{\nano\meter} wide 
segments and performed numerical fast Fourier transformations on each of them individually. Afterwards we 
matched the frequency amplitude ranges by normalizing the different segments to each other, colour coded the 
resulting profiles, and merged the different wavelength ranges to obtain the plot shown in Figure \ref{fig:segfu}. The 
different Fourier components of the resonator's whispering gallery modes (WGMs) are clearly visible as bands evenly spaced by an integer 
multiple of the lowest observable base frequency. This frequency component is directly related to 1/FSR and 
corresponds well with the theoretical estimated value based on the size of the discs. The black dotted lines 
are fits to this value and its different higher-order multiples.

\clearpage

\section*{Tuning of the Resonators}

\begin{figure*}[hb]
  \includegraphics[width=\textwidth]{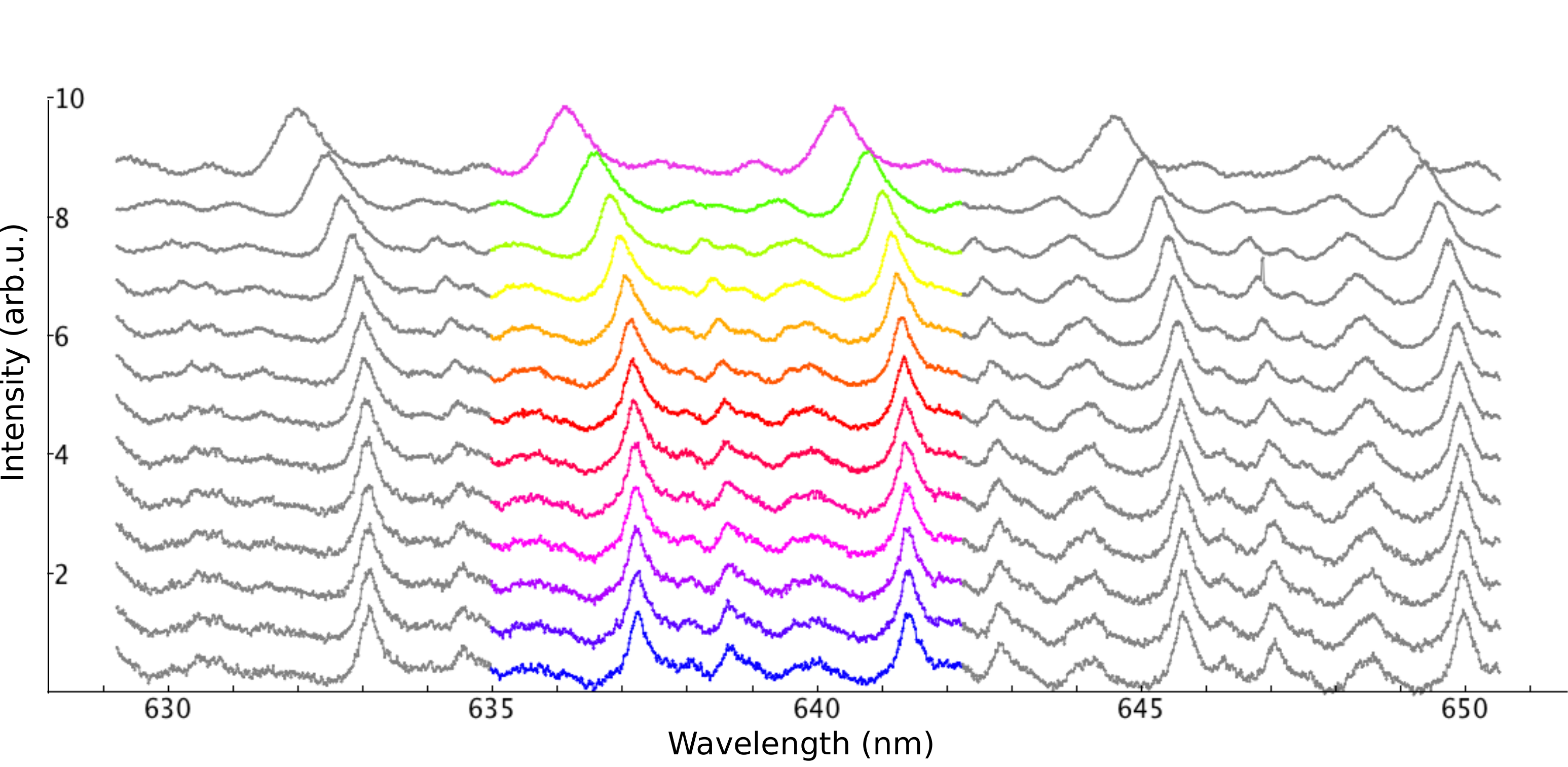}
  \caption{Permanent tuning of a resonator under vacuum}  
  \label{fig:tuning}
\end{figure*}

Figure \ref{fig:tuning} shows the tuning of a resonator under vacuum conditions. To tune the resonator, a \unit{405}{\nano\meter}
laser was focussed on its rim. The spectra shown were acquired from bottom to top in equal timesteps of \unit{30}{\second}. 
After some steps the mode begins to shift.
Laser induced shifts of the resonance of resonators were previously observed by us~\cite{Wolters2010_} and others~\cite{Others}.
Its physical or chemical origin is not understood yet, but it is assumed to be caused by a modification
of the resonator's local shape or a change of the index of refraction of the resonators material.
It is worth noting that tuning of the resonator is possible even without oxygen present.

This is a very important feature, since it enables tuning of the resonator's whispering gallery modes to the zero phonon lines of 
the nitrogen vacancy centres. Since no oxygen is needed for this procedure, it can be used in a cryostat at low temperatures. 
In addition to the possibility of tuning the resonances to lines of the optical emitters, it would be also possible to match 
resonances of different resonators and achieve controlled coupling among them.

\clearpage

\section*{Background correction}

In the case of continuous excitation of an emitter with intensity $<n_{a}>=a$, second order correlation 
functions $g^{(2)}_{a}$ and the quantum mechanically uncorrelated background of the intensity 
$<n_{b}>= b$ with  $g^{(2)}_{b}=1$, the joint second order correlation function is given by:
\begin{eqnarray}
 g^{(2)}_{ab}&=& \frac{a^{2}g^{(2)}_{a} + b^{2}+2 a b}{(a+b)^{2}}.
 \label{eq:g2cw}
\end{eqnarray}
By knowing the intensities $a$ and $b$ it one can 
calculate the $g^{(2)}_{a}$ function of the bare emitter $a$ from a measurement of the joint correlation function $g^{(2)}_{ab}$.

In case of pulsed excitation this background correction is more complicated, since the emission of emitter and background 
are correlated by the excitation laser. This problem can be avoided by only looking at time intervals corresponding 
to one laser period. In the analysis of an experiment's $g^{(2)}$-data acquired with a Hanbury Brown and Twiss setup, 
this can be achieved by binning together all events in the time windows $T_{\frac{max}{min}}=n \Delta T \pm \frac{\Delta T}{2}$, with the laser 
repetition time $\Delta T$ ($=$ \unit{25}{\nano\second} in our experiment) and $n$ being an integer.

\begin{figure*}
  \includegraphics[width=\textwidth]{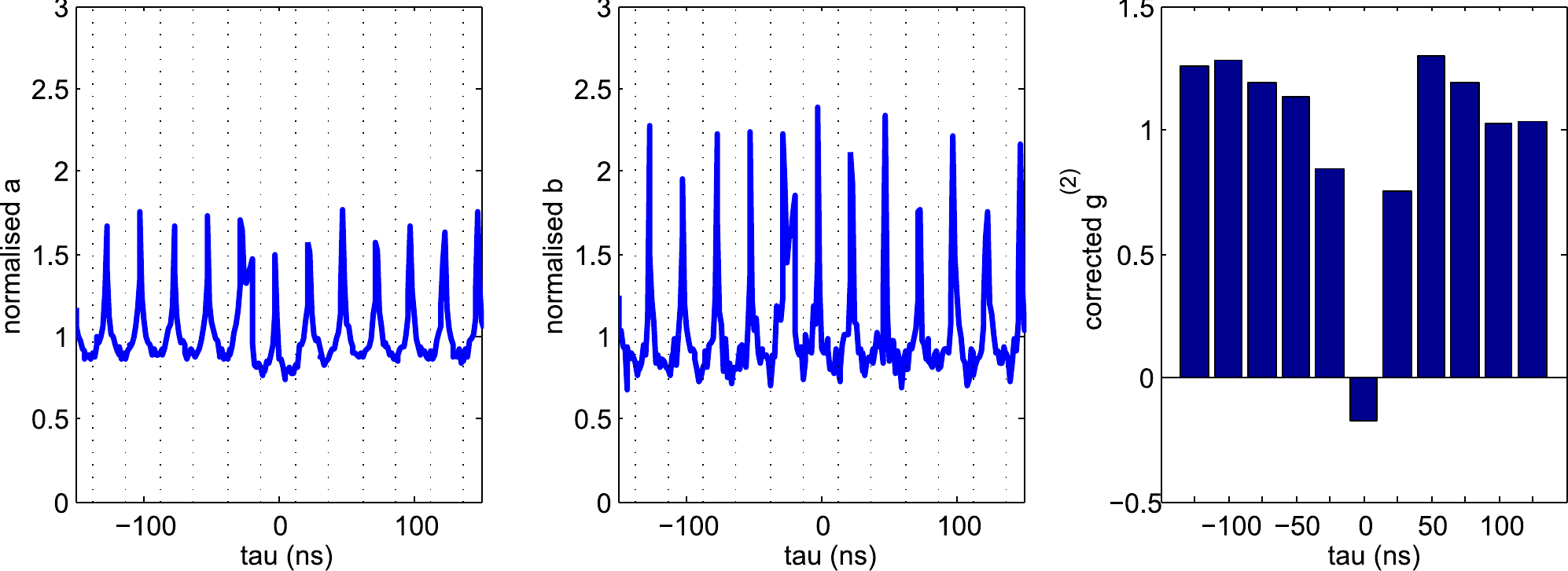}
  \caption{Data used for the background correction. a and b are normalised coincidence counts
  measured with pulsed excitation at the location of an NV centre at the resonator's rim (in a)) and
  at a location without an NV center next to it also on the rim (in b)),
  respectively. Both were normalised to an average of one at long times $>$\unit{1.3}{\micro\second} over 256 periods). A binsize of \unit{1.28}{\nano\second}
  was used. The dotted lines show the bins, as they were used to obtain c. c shows the resulting antibunching with a 
  $g^{(2)}(0)=-0.18\pm0.21$.}  
  \label{fig:ab_correction}
\end{figure*}

Figure~\ref{fig:ab_correction} a and b show the normalised coincidence counts
measured with pulsed excitation at the location of an NV centre at the resonator's rim (in a)) and
at a location without an NV center next to it also on the rim (in b)),
respectively.  Thus the data in figure~\ref{fig:ab_correction}a) corresponds to coincidences from
signal and background whereas figure~\ref{fig:ab_correction}b) represents the background coincidences only.
Binning to intervals of \unit{1.28}{\nano\second} was performed for better visibility.
Normalisation was done by averaging for long times ($>$\unit{1.3}{\micro\second} over 256 periods), where contributions from 
the NV centre's bunching behaviour are negligible. 
Also the photons normally missed in an HBT like start-stop configuration were taken into account.
A comparison of the background photons rate with the intensity of the combined signal gave $a=\frac{1}{3}$ and $b=\frac{2}{3}$, equivalent 
to a background to signal ratio of 2.
Figure~\ref{fig:ab_correction} c shows the resulting corrected $g^{(2)}(\tau)$ function, with bins sizes corresponding to the 
laser repetition time.

For zero time delay $g^{(2)}(0)=-0.18$. The error bar for this value stems from three contribution, with the main contribution stemming from the 
uncertainties in the ratio $\frac{b}{a}$, which we assumed to be 10~\% ($\frac{b}{a}=2.0\pm0.2$). This error is due to the fact that 
background and signal were measured at neighboring positions on the resonator's rim, but not at the same one (see Figure 5 b of the main text,
$\Delta(\frac{a}{b})=0.20$).
The other two contributions are the statistical variations in the photon number for the zero time delay bin, which were both assumed to be Possonian
($\Delta(a)=0.01$ and $\Delta(b)=0.07$).
This results in an error of $\Delta(g^{(2)}(0))=0.21$ after adding the contributions 
via Gaussian error propagation.

\end{document}